\documentclass[letterpaper, 10 pt, conference]{ieeeconf}  

\IEEEoverridecommandlockouts                              
\title{\LARGE \bf
Safe Control of Multi-Agent Systems with Minimal Communication

}

\author{Mo Yang, Jing Yu and Necmiye Ozay
\thanks{This work was supported in part by ONR CLEVR-AI MURI (\#N00014-21-1-2431). MY and NO are with the University of Michigan, Ann Arbor, MI, USA (emails:{\tt \{sprkyang,necmiye\}@umich.edu}). JY is with the University of Washington, Seattle, WA, USA (email:{\tt jing5@uw.edu}). The code for this paper can be found at \url{https://github.com/Nephren17/multiagent_lowrankctrl}.} }

\usepackage{url}
\usepackage{graphicx}

\usepackage{amsthm}
\usepackage{amsmath}
\usepackage{amssymb}
\usepackage{amsfonts}
\usepackage{cite}

\usepackage{amsmath}
\usepackage[dvipsnames]{xcolor}

\usepackage[colorlinks=true, linkcolor=blue]{hyperref}
\usepackage{cleveref}

\usepackage{algorithm}
\usepackage{algorithmicx}
\usepackage{algpseudocode}
\usepackage{booktabs}
\usepackage{multirow}
\usepackage{subcaption}

\newtheorem{proposition}{Proposition}[]
\newtheorem{lemma}{Lemma}[]
\newtheorem{remark}{Remark}
\newtheorem{problem}{Problem}

\crefname{assumption}{assumption}{assumptions}
\Crefname{assumption}{Assumption}{Assumptions}

\newcommand{\mb}[1]{\mathbf{#1}}

\newcommand{\N}{\mathcal{N}}
\newcommand{\Ob}{\mathcal{O}}
\newcommand{\D}{\mathcal{D}}
\newcommand{\U}{\mathcal{U}}
\newcommand{\Lo}{\mathcal{L}}

\newtheorem{constraint}{Constraint}

\crefname{constraint}{Constraint}{Constraints}
\Crefname{constraint}{Constraint}{Constraints}

\newcommand{\blkdiag}{\text{blkdiag}}
\newcommand{\bphi}[1]{\mathbf{\Phi}^{#1}}
\newcommand{\spar}[1]{\text{sp}\left(#1\right)}

\newcommand{\baseline}{\textsc{Baseline}}
\newcommand{\decentral}{\textsc{Decentral}}

\begin{document}

\maketitle
\thispagestyle{empty}
\pagestyle{empty}

\begin{abstract}

In many multi-agent systems, communication is limited by bandwidth, latency, and energy constraints. Designing controllers that achieve coordination and safety with minimal communication is critical for scalable and reliable deployment. 
This paper presents a method for designing controllers that minimize inter-agent communication in multi-agent systems while satisfying safety and coordination requirements, while conforming to communication delay constraints. The control synthesis problem is cast as a rank minimization problem, where a convex relaxation is obtained via system level synthesis. 
Simulation results on various tasks, including trajectory tracking with relative and heterogeneous sensing, demonstrate that the proposed method significantly reduces inter-agent transmission compared to baseline approaches.
\end{abstract}


\section{Introduction}

Multi-agent systems, such as robotic swarms, distributed sensor networks, and autonomous vehicle fleets, are increasingly deployed to perform complex tasks in dynamic and uncertain environments. 
These systems are often composed of individual agents with decoupled dynamics and limited, heterogeneous sensing capabilities. For instance, fleets of vehicles may only have access to relative position measurements with respect to nearby neighbors, rather than absolute coordinates \cite{marshall2024convex}. Therefore, agents must rely on communication to share information and coordinate actions. However, communication in such systems is often constrained by factors such as latency, bandwidth limitations, and network topology. 
For example, in space exploration, planetary rovers must share information about terrain and obstacles with minimal communication to preserve energy \cite{bajracharya2008autonomy, Fang2024Space}.
Therefore, designing control algorithms that enable multi-agent systems to accomplish prescribed tasks while minimizing communication is a key challenge \cite{zuo2022fish}.

There has been extensive research on communication-efficient control in networked systems. Event-triggered control has been widely used to reduce unnecessary transmissions by only communicating when certain state-dependent conditions are met \cite{heemels2012introduction,kia2015distributed, nowzari2019event}. If the communication structure and other constraints of the network are pre-defined, distributed control can be used to design structured controllers \cite{ lamperski2015optimal, rotkowitz2005characterization, wang2019system,yu2021localized,zheng2020equivalence}. Another approach is to develop
low-rank approximations and sparsity-promoting controllers that result in sparse communication among agents \cite{wu2017sparsity, lin2013design, molzahn2017survey, aspeel2023low}. 

More recently, a line of work has investigated control designs that minimize the number of messages transmitted from the sensors to the actuators within a single agent system for finite-horizon control problems \cite{aspeel2023low, aspeel2024minimal}. This approach formulates sensor-to-actuator transmission optimization as a rank minimization problem, enabling the design of controllers that are guaranteed to achieve safe control tasks with the minimal number of sensor-to-actuator messages. 

Inspired by \cite{aspeel2023low}, in this work, we formulate the multi-agent minimal communication control problem and propose an algorithm based on rank minimization and system level synthesis \cite{anderson2019system} to compute a distributed controller. Our method naturally handles communication delays in the network. The numerical experiments show that the proposed controller significantly reduces the number of messages sent among agents for safe control tasks compared to the benchmarks. 

\textbf{Notation. } The set of positive integers is denoted as $\mathbb{N}_+$. We denote by $I$ the identity matrix, with dimension determined by context unless explicitly stated otherwise. The notation $\blkdiag(A_1\, \ldots,\, A_n)$ denotes a diagonal block matrix with diagonal blocks $A_1,\, \ldots,\, A_n$. Given an index set $\N$, we write $\N_{-i}$ to mean the set excluding the element $i$, i.e., $\N \setminus \{i\}$. For a matrix $\mathbf{M}\in \mathbb{R}^{m\times n}$, $M(i,j)$ denotes its element in the $i$th row and the $j$th column.

\section{Problem Statement}

Consider a network of $N$ agents indexed by the set $\N := \{1,\,2,\,\ldots,\,N\}$. Each agent $i$ has decoupled linear time-varying dynamics with coupled measurements
\begin{equation}
    \label{eq: system dynamics}
    \begin{aligned}
        x^i_{t+1} &=  [A]^{ii}_t x^i_t + [B]^{ii}_t u^i_t + w^i_t, \\
        y^i_{t} &= \sum_{k\in\Ob_i}[C]^{ik}_t x^k_t + v^i_t,
    \end{aligned}
\end{equation}
over a finite horizon $T$, where $x^i_t \in \mathbb{R}^{n_x^i}$, $u^i_t \in \mathbb{R}^{n^i_u}$, $w^i_t \in \mathbb{R}^{n^i_x}$, $y^i_t \in \mathbb{R}^{n^i_y}$ and $v^i_t \in \mathbb{R}^{n^i_y}$ represent local state vector, control action, state disturbances, measurement, and measurement noise, respectively. The state dynamics of each agent is decoupled and time-varying, driven by $[A]^{ii}_t$ and $[B]^{ii}_t$. The set $\mathcal{O}_i$ contains the indices associated with the agents that influence the measurements of agent $i$.

We use $x_t \in \mathbb{R}^{n_x}$, $u_t \in \mathbb{R}^{n_u}$, $w_t \in \mathbb{R}^{n_x}$, $y_t \in \mathbb{R}^{n_y}$ and $v_t \in \mathbb{R}^{n_y}$ to denote the joint vectors of the $N$ agents with
$
x_t = \begin{bmatrix} \left(x_t^1\right)^\top; \left(x_t^2\right)^\top; \cdots; \left(x_t^N\right)^\top \end{bmatrix}^\top$. The concatenated vectors $u_t$, $w_t$, $y_t$, $v_t$ are defined similarly. We define the global dynamics matrices at time $t$, denoted by $A_t$, $B_t$, and $C_t$, by assembling the submatrices $[A]^{ij}_t$, $[B]^{ij}_t$, and $[C]^{ij}_t$ into block matrices. Specifically, each $ [\,\cdot\,]_t^{ij} $ is placed in the \( (i, j) \)th block of the corresponding global matrix.

In this paper, we assume the coordination task and the safety constraints for the agents can be collectively expressed as constraint sets. In particular, we consider polyhedra $\mathcal{X}_t \subset \mathbb{R}^{n_x}$, $\mathcal{W}_t\subset \mathbb{R}^{n_x}$, $\mathcal{U}_t\subset \mathbb{R}^{n_u}$, and $\mathcal{V}_t\subset \mathbb{R}^{n_y}$. 
\begin{constraint}[Safety]
\label{const:safety}
    For all $x_0\in \mathcal{X}_0$, $w_t \in \mathcal{W}_t$, and $v_t \in \mathcal{V}_t$, ensure that $u_t\in \mathcal{U}_t$ and $x_t\in \mathcal{X}_t$ for all $t\leq T$. 
\end{constraint}

To maintain safety, each agent is equipped with a local controller. The local control action for each agent $i$ at time $t$ is computed using a combination of its previously collected local measurements and the measurements transmitted from other agents $j$:
\begin{equation}
\label{eq:local_controller}
    u^i_t = \sum_{\tau \leq t} [K]^{ii}_{(t,\tau)} y^i_\tau + \sum_{j\in\N_{-i}}  \underbrace{\sum_{\tau \leq t}  [K]^{ij}_{(t,\tau)}y^j_\tau}_{[u^i_t]_j} \,, 
\end{equation}
where $[K]^{ii}_{(t,\tau)} \in \mathbb{R}^{n^i_u\times n^i_y}$ and $[K]^{ij}_{(t,\tau)} \in \mathbb{R}^{n^i_u\times n^j_y}$ for all $\tau = 0,1,\ldots, T$. We will use $[u^i_t]_j \in \mathbb{R}^{n^i_u}$ to denote $\sum_{\tau \leq t}  [K]^{ij}_{(t,\tau)} y^j_\tau$, i.e., the contribution of agent $j$'s information to the computation of the control action of agent $i$ at time $t$. Moreover, we denote the concatenation of $[K]^{ij}_{(t,\tau)}$ over $t,\,\tau \in \{0,\, \ldots,\,T\}$ for fixed $i,\,j \in \N$ as
\begin{equation}
\label{eq:K_ij}
\mb{K}^{ij} = \begin{bmatrix}
[K]^{ij}_{(0,0)} & 0 & \ldots & 0\\
[K]^{ij}_{(1,0)} & [K]^{ij}_{(1,1)} & 0 &\vdots \\
\vdots &   & \makebox[1cm][r]{$\ddots$} & 0\\
[K]^{ij}_{(T,0)} & \cdots & [K]^{ij}_{(T,T-1)} & [K]^{ij}_{(T,T)}
\end{bmatrix},
\end{equation}
where $\mb{K}^{ij} \in \mathbb{R}^{(T+1)n^i_u \times(T+1)n^j_y}$ is block 
 lower-triangular. 

Networked systems often experience communication delays due to the limited bandwidth of communication channels. In this paper, we will model the communication delay from agent $j$ to $i$ as a fixed worst-case latency $\ell^{ij} \in \mathbb{N}_+$, meaning that at time $t$, agent $i$ has access to agent $j$’s information only up to time $t - \ell^{ij}$. This delay imposes additional sparsity constraints on the local controllers, as detailed below. 
\begin{constraint}[Communication delay]
\label{const:delay}
For all $i,\, j\in \N$ and $i \not = j$, $[K]^{ij}_{(t,\tau)} = 0$ if $t-\tau \leq \ell^{ij}-1$.
\end{constraint}
We will assume that the communication delay constraints satisfy the quadratic invariance (QI) property \cite{rotkowitz2005characterization,lessard2011quadratic}, which guarantees that any such constraints on the controller can be translated to 
convex constraints after controller reparameterization, enabling convex reformulation of optimal constrained control design via approaches such as Youla reparameterization \cite{skaf2010design} and system level synthesis \cite{wang2019system}.  In practice, QI is often satisfied in settings where the communication pattern mirrors the physical dynamics coupling, such as vehicle platoons and power grid frequency regulation.

\begin{figure}[h]
    \centering
    \includegraphics[width=0.85\linewidth]{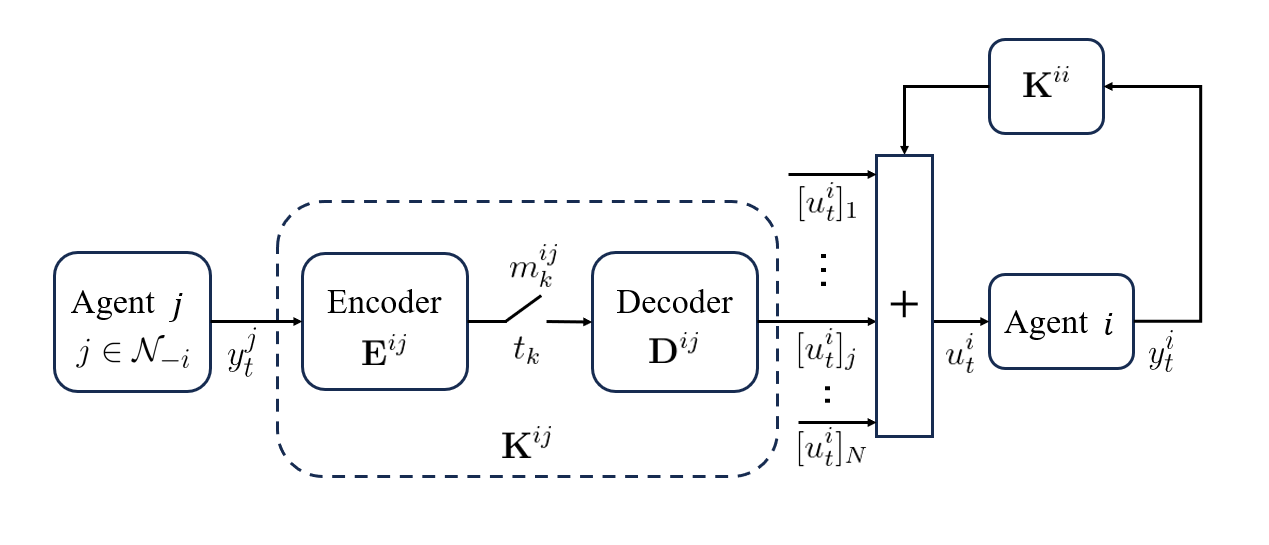}
    \caption{Local controller structure with encoder and decoder where $\mathbf{D}^{ij}$ and $\mb{E}^{ij}$ are defined in \eqref{eq:DE}.}
    \label{fig: encode decode diagram}
\end{figure}

Note that without further structural constraints on the gains $[K]^{ij}_{(t,\tau)}$, the implementation of \eqref{eq:local_controller} essentially requires each agent to transmit their measurements to all other agents at every time step $t$. However, in applications where communication across agents is costly or difficult, such as underwater vehicles, the number of inter-agent transmissions should be minimized. Moreover, some coordination tasks inherently do not require frequent inter-agent communication to be effectively executed. Motivated by \cite{aspeel2023low} where the authors decompose the controller with an encoder-decoder structure in order to minimize sensor-to-actuator message transmissions in the single-agent setting, we similarly decompose the computation of $[u^i_t]_j$ for each $j \in \N_{-i}$ in \eqref{eq:local_controller} with the following structure: for $0 \leq t_1 \leq t_2 \leq \ldots \leq t_{r^{ij}}\leq t$ with $r^{ij} \in \mathbb{N}_+$, let
\begin{equation}
\label{eq:local_DE}
\begin{aligned}
m^{ij}_k &= \sum_{\tau \leq t_k} \left(e^{ij}_{(k,\tau)} \right)^\top y^j_{\tau} \\
[u^i_t]_j &= \sum_{k\text{ s.t. }t_k \leq t} d^{ij}_{(t,k)} m^{ij}_k
\end{aligned}
\end{equation}
where \scalebox{0.8}{$e^{ij}_{(k,\tau)} \in \mathbb{R}^{n^j_y}$} serves as a linear \textbf{encoder} at agent $j$, and each $m^{ij}_k \in \mathbb{R}$ is an encoded message sent from agent $j$ to agent $i$ at time $t_k$. The number of messages sent from $j$ to $i$ is $r^{ij}$. The contribution of agent $j$'s information to $u^i_t$ is then computed by using a local \textbf{decoder} $d^{ij}_{(t,k)}\in\mathbb{R}^{n^i_u}$ at agent $i$. This controller structure is illustrated in \Cref{fig: encode decode diagram}.

Decomposition \eqref{eq:local_DE} can be interpreted as a low-rank factoring $\mb{K}^{ij} = \mb{D}^{ij} \mb{E}^{ij} $ where 
\begin{equation}
\label{eq:DE}
\begin{aligned}
    \mathbf{D}^{ij} &:= 
\begin{bmatrix}
d^{\,ij}_{(0,1)} & \cdots & d^{\,ij}_{(0,r^{ij})} \\
\vdots &  & \vdots \\
d^{\,ij}_{(T,1)} & \cdots & d^{\,ij}_{(T,r^{ij})}
\end{bmatrix} \in \mathbb{R}^{(T+1)n^i_u \times r^{ij}}\\
\mathbf{E}^{ij} &:=
\begin{bmatrix}
e^{\,ij \,\,\,\,\,\top}_{(1,0)}& \cdots & e^{\,ij \,\,\,\,\,\,\top}_{(1,T)} \\
\vdots &  & \vdots \\
e^{\,ij \,\,\,\,\,\,\,\top}_{(r^{ij},0)} & \cdots & e^{\,ij \,\,\,\,\,\,\,\,\,\top}_{(r^{ij},T)}
\end{bmatrix} \in \mathbb{R}^{r^{ij} \times (T+1) n^j_y}.
\end{aligned}
\end{equation}


A key insight from Aspeel et al. \cite{aspeel2023low} is that 
the matrix $\mb{K}^{ij}$ always admits such a factorization that provides a causal encoder $\left\{e^{ij}_{(k,\tau)}\right\}_{k\in [r^{ij}]}^{\tau = 0,\ldots,t_k}$, causal decoder $\left\{d^{ij}_{(k,\tau)}\right\}_{k\in [r^{ij}]}^{\tau = 0,\ldots,t_k}$, and the corresponding message sending times $\{t_k\}_{k\in[r^{ij}]}$ (with a total number of $r^{ij}$ messages), which can be used to construct local controllers \eqref{eq:local_DE}. In particular, given a matrix $\mb{K}^{ij}$, the message sending times, encoder, and decoder can be computed with Algorithm 1 of \cite{aspeel2023low}. 
\begin{lemma}[{\cite[Theorem 1]{aspeel2023low}}]
\label{thrm:aspeel}
For all $i,\, j \in \N$, there exists a causal factorization of matrix $\mathbf{K}^{ij}$ such that $\mathbf{K}^{ij} =\mb{D}^{ij}\mb{E}^{ij}$ 
with $r^{ij} = \text{rank}\left(\mb{K}^{ij}\right)$. Moreover, there exists $\{t_k\}_{k\in[r^{ij}]}$ such that $0 \leq t_1 \leq t_2 \leq \ldots \leq t_{r^{ij}} \leq T$, and it holds that $e^{ij}_{(k,\tau)} = 0 $ whenever $\tau > t_k$ and $d^{ij}_{(t,k)} = 0$ whenever $t_k > t$.
\end{lemma}

 Based on \eqref{eq:DE}, problem statement of this paper is as follows.
\begin{problem}
\label{prob:1}
Given the dynamics in \eqref{eq: system dynamics} and the controller structure in \eqref{eq:local_DE}, find message  transmission times $\{t_k\}_{k\in[r^{ij}]}$, corresponding encoder-decoder matrices $\left\{e^{ij}_{(k,\tau)}\right\}_{k\in [r^{ij}]}^{\tau = 0,\ldots,t_k}$, and $\left\{d^{ij}_{(k,\tau)}\right\}_{k\in [r^{ij}]}^{\tau = 0,\ldots,t_k}$ for all agents $i,\,j\in\N$ such that \Cref{const:safety} and \Cref{const:delay} are satisfied for all initial conditions, disturbance and noise realizations, and the total  number of inter-agent messages, i.e., $\sum_{i,\,j \in\N}r^{ij}$, is minimized.
\end{problem}
\vspace{-0.2in}

To address \Cref{prob:1}, we now generalize the approach of \cite{aspeel2023low} to the networked system setting by transforming \Cref{prob:1} into an equivalent rank minimization problem over the local gains $[K]^{ij}_{(t,\tau)}$. This is enabled by the following result that relates the number of inter-agent messages required by \eqref{eq:local_DE} and the rank of $\mathbf{K}^{ij}$.
\begin{proposition}
\label{lem:equivalent}
The optimal solution to \Cref{prob:1}, i.e., the minimum number of inter-agent messages, is equal to the minimum of the following optimization problem:
\begin{equation}
\label{eq:rank_min}
\begin{aligned}
&\min_{\mb{K}^{ij}} &&\sum_{i,\,j\in\N, \, i\not= j} \text{rank}\left(\mb{K}^{ij}\right) \\
&\text{s.t.} &&\text{\eqref{eq: system dynamics}, \eqref{eq:local_controller}, \Cref{const:safety} and \ref{const:delay}.}
\end{aligned}
\end{equation}
\end{proposition}
\textit{Proof: }
By \Cref{thrm:aspeel}, there exists a causal factorization for any $\mb{K}^{ij}$ with corresponding $r^{ij}$ message sending times. Note that $r^{ij}$ is the number of messages agent $j$ sends to agent $i$ using controller \eqref{eq:local_controller} implemented with $\mb{K}^{ij}$ and $\text{rank}\left(\mb{K}^{ij}\right) = r^{ij}$. Therefore, minimizing the total number of inter-agent messages is equivalent to minimizing the sum of rank of $\mb{K}^{ij}$ over all $i,\,j\in \N$ with $i \not= j$.
\hfill $\square$

Thus, \Cref{prob:1} is effectively reduced to the rank minimization problem \eqref{eq:rank_min}.

\section{Convex relaxation via SLS}

We note that even if we replace the rank minimization objective function in \eqref{eq:rank_min} with a convex envelope,  \Cref{const:safety} and \Cref{const:delay} still make \eqref{eq:rank_min} non-convex. 
In particular, it is well established that general sparsity constraints, which capture delays and other communication constraints in distributed controllers, lead to non-convex constrained control problems. When the sparsity constraints satisfy quadratic invariance \cite{rotkowitz2005characterization}, the constrained control problem can be cast as an equivalent convex program through a reparameterization of the controller, e.g., Youla parameterization, input-output parameterization \cite{zheng2021sample}, and SLS \cite{wang2019system}. Furthermore, it has been shown that \Cref{const:safety} can be expressed equivalently as linear constraints in terms of the SLS parameter \cite{chen2019system}. 
Therefore, in this section, we leverage SLS to perform convex relaxation of \eqref{eq:rank_min}. 

\subsection{System level synthesis (SLS)}
We now give a brief introduction to SLS.
 Define notation $\mathbf{x} := \begin{bmatrix} {x}_0^\top & \cdots & {x}_T^\top \end{bmatrix}^\top,\,\,
\mathbf{u} := \begin{bmatrix} {u}_0^\top & \cdots & {u}_T^\top \end{bmatrix}^\top,\,\,
\mathbf{y} := \scalebox{0.7}{$\begin{bmatrix} {y}_0^\top & \cdots & {y}_T^\top \end{bmatrix}^\top$},\,\,
\mathbf{w} := \scalebox{0.7}{$\begin{bmatrix} {x}_0^\top & {w}_0^\top & \cdots & {w}_{T-1}^\top \end{bmatrix}^\top$}, \,\,
\mathbf{v} := \begin{bmatrix} {v}_0^\top & \cdots & {v}_T^\top \end{bmatrix}^\top$, $\mathcal{A} := \blkdiag(A_0,\, \ldots,\,A_{T-1},\, 0)$, $\mathcal{B} := \blkdiag(B_0,\,\ldots,\,B_{T-1},\,0)$, $\mathcal{C} := \blkdiag(C_0,\,\ldots,\, C_{T})$, and let $\mathcal{Z}$ be the block-downshift operator. Then the global dynamics of \eqref{eq: system dynamics} can be written as 
\begin{equation}
\label{eq:global}
    \begin{aligned}
   \mb{x} &= \mathcal{Z}\mathcal{A}\mb{x} + \mathcal{Z}\mathcal{B}\mb{u} + \mb{w} \\
\mb{y} &= \mathcal{C}\mb{x} + \mb{v} \,.  
\end{aligned}
\end{equation}

SLS parameterizes all achievable closed-loop responses of \eqref{eq:global} under a linear output feedback control law $\mb{u} = \mb{K}\mb{x}$ where $\mathbf{K} \in \mathbb{R}^{ (T+1)n_u \times (T+1 ) n_y}$ and 
\begin{equation}
    \mb{K} :=  \begin{bmatrix}
K_{(0,0)} \\
K_{(1,0)} & K_{(1,1)} \\
\vdots & \ddots & \ddots \\
K_{(T,0)} & \cdots & K_{(T,T-1)} & K_{(T,T)}
\end{bmatrix} .
\label{eq:K}
\end{equation}
 Note that this is a centralized description for the local controllers \eqref{eq:local_controller} (equivalently \eqref{eq:K_ij}). For all $t,\,\tau$, the sub-block $K_{(t,\tau)}$ is itself a block matrix, whose $(i,j)$th mini-block corresponds to $[K]^{ij}_{(t,\tau)}$ in \eqref{eq:K_ij}.
Now consider 
\begin{equation}
    \label{eq:clm}
    \begin{bmatrix}
        \mb{x} \\  \mb{u}
    \end{bmatrix} = 
        \begin{bmatrix}
        \bphi{xx} & \bphi{xy} \\  \bphi{ux} &\bphi{uy}
    \end{bmatrix}
    \begin{bmatrix}
        \mb{w} \\  \mb{v}
    \end{bmatrix}
\end{equation}
where $\bphi{} := (\bphi{xx},\, \bphi{xy} ,\, \bphi{ux} ,\, \bphi{uy})$ is the closed-loop response that maps exogenous disturbances $\mb{w}$ and measurement noise $\mb{v}$ to $\mb{x}$ and $\mb{u}$ in the closed loop under controller $\mb{K}$. In particular, $\bphi{xx} = \left({I} - \mathcal{Z} \mathcal{A} - \mathcal{Z} \mathcal{B} \mathbf{K} \mathcal{C} \right)^{-1}$, 
$\bphi{xy} = \bphi{xx} \mathcal{Z} \mathcal{B} \mathbf{K}$, 
$\bphi{ux} = \mathbf{K} \mathcal{C} \bphi{xx}$, 
$\bphi{uy} = \mathbf{K} + \mathbf{K} \mathcal{C} \bphi{xx} \mathcal{Z} \mathcal{B} \mathbf{K}$, where $\bphi{xx}$, $\bphi{ux}$, and $\bphi{uy}$ are block lower triangular matrices while $\bphi{xy}$ is strictly block lower triangular matrices. 

SLS provides an affine characterization of the space of all achievable $\bphi{}$ and a direct relationship between $\bphi{}$ and $\mb{K}$ in the following result.
\begin{lemma}[{\cite[Lemma 1]{hassaan2022system}}]
   For system \eqref{eq:global}, the following are true:
\begin{enumerate}
    \item The affine subspace defined by
    \begin{equation}
    \label{eq:sls}
    \begin{aligned}
    \begin{bmatrix}
        {I} - \mathcal{Z}\mathcal{A} & -\mathcal{Z}\mathcal{B}
    \end{bmatrix}
    \begin{bmatrix}
        \bphi{xx} & \bphi{xy} \\
        \bphi{ux} & \bphi{uy}
    \end{bmatrix}
    &= 
    \begin{bmatrix}
        I & 0
    \end{bmatrix} \\
    \begin{bmatrix}
        \bphi{xx} & \bphi{xy} \\
        \bphi{ux} & \bphi{uy}
    \end{bmatrix}
    \begin{bmatrix}
        {I} - \mathcal{Z}\mathcal{A} \\
        -\mathcal{C}
    \end{bmatrix}
    &= 
    \begin{bmatrix}
        I \\
        0
    \end{bmatrix} 
    \end{aligned}
    \end{equation}
    parameterizes all possible closed-loop responses \eqref{eq:clm} under a linear output feedback controller $\mb{u} = \mb{K}\mb{x}$.
    
    \item For any block lower triangular matrices $\bphi{xx}, \bphi{xy}, \bphi{ux}, \bphi{uy}$ satisfying \eqref{eq:sls}, the controller 
    \begin{equation}
    \label{eq:K-phi}
    \mathbf{K} = \bphi{uy} - \bphi{ux} \left(\bphi{xx}\right)^{-1} \bphi{xy}
    \end{equation}
    achieves the desired closed-loop system response \eqref{eq:clm}.
\end{enumerate}
\end{lemma}

\subsection{Reformulation via SLS}
As discussed in \cite{aspeel2023low, chen2019system}, the safety Constraint \ref{const:safety} can be expressed equivalently as convex constraints in terms of $\bphi{}$. Moreover, the control problem \eqref{eq:rank_min} with the communication delay Constraint \ref{const:delay} can be expressed as an equivalent convex optimization problem in $\bphi{}$ \cite{anderson2019system}. Therefore, we aim to express and solve \eqref{eq:rank_min} in terms of $\bphi{}$ by searching over all the $\bphi{}$ that satisfy \eqref{eq:sls} and Constraints \ref{const:safety} and \ref{const:delay}. By computing the global controller $\mb{K}$ from $\bphi{}$ using \eqref{eq:K-phi}, the local controller $\mb{K}^{ij}$ can then be identified by setting $[K]^{ij}_{(\tau,t)}$ as the $(i,j)$th mini-block of sub-blocks $K_{(\tau,t)}$ from $\mathbf{K}$ for all $t \text{, }\tau \in \{0,\,1,\, \ldots, T\}$. 

However, due to the nonlinear relationship between $\mb{K}$ and $\bphi{}$ in \eqref{eq:K-phi}, it is unclear how the objective function of \eqref{eq:rank_min} should be expressed in terms of $\bphi{}$. 
We now present a result that upper bounds the rank of $\mathbf{K}^{ij}$ with the rank of $\bphi{}$, enabling a convex relaxation of \eqref{eq:rank_min} via SLS. 

Following the convention of \eqref{eq:K}, we denote the $(t,\tau)$th sub-block matrix of $\bphi{xx}$ as $\Phi^{xx}_{(t,\tau)}\in \mathbb{R}^{n_x\times n_x}$ for $t,\,\tau \in \{1,\,\ldots,\, T\}$ with $\Phi^{xx}_{(t,\tau)}=0$ for all $\tau > t$ and analogously for $\Phi^{xy}_{(t,\tau)}$, $\Phi^{ux}_{(t,\tau)}$, and $\Phi^{uy}_{(t,\tau)}$. Furthermore, we will denote the $(i,j)$th $n^i_x $ by $ n^j_x$ mini-block matrix of $\Phi^{xx}_{(t,\tau)}$ as $\Phi^{xx,\,ij}_{(t,\tau)}$ for $i,\,j \in \N$ and analogously for $\Phi^{xy,\,ij}_{(t,\tau)}$, $\Phi^{ux,\,ij}_{(t,\tau)}$, and $\Phi^{uy,\,ij}_{(t,\tau)}$. Concatenating all $\Phi^{xx,\,ij}_{(t,\tau)}$ similar to \eqref{eq:K_ij}, we denote the final block lower triangular matrix $\bphi{xx,\,ij} \in \mathbb{R}^{ (T+1)n^i_x \times (T+1)n^j_x }$, and define $\bphi{xy,\,ij} \in \mathbb{R}^{ (T+1)n^i_x \times (T+1)n^j_y }$, $\bphi{ux,\,ij} \in \mathbb{R}^{ (T+1)n^i_u \times (T+1)n^j_x }$, and $\bphi{uy,\,ij} \in \mathbb{R}^{ (T+1)n^i_u \times (T+1)n^j_y }$ similarly. 
For simplicity, we will focus on the case of $N=2$ agents in the following. 

\begin{proposition} 
\label{lem:phi-rank}
For any block lower triangular matrices $\bphi{xx}, \bphi{xy}, \bphi{ux}, \bphi{uy}$ satisfying \eqref{eq:sls}, if for all $\tau,\,t \in \{ 1,\, \ldots, T\}$, $\Phi^{xx,\,ij}_{(t,\tau)} = 0$ either for all $i>j$ or for all $i<j$, then the corresponding controller $\mathbf{K}$ computed with \eqref{eq:K-phi} satisfies that for $i,j \in \{1,2\}$ and $i\not = j$,
\begin{equation}
\label{eq:sls-rank}
\text{rank}(\mb{K}^{ij}) \leq \!\!\underset{*\in \{xx, xy,ux,uy\}}{\sum} \!\! \text{rank}(\bphi{*,\,ij}).
\end{equation}
\end{proposition}
The proof can be found in the Appendix.
With \Cref{lem:phi-rank}, \eqref{eq:rank_min} can be heuristically solved with a convex relaxation via the SLS parameterization. Instead of directly minimizing the number of messages, we minimize its upper bound. Moreover, we will use a surrogate objective function by replacing the rank term in the upper bound from \Cref{lem:phi-rank} with its reweighted nuclear norm. The resulting SLS controller will then satisfy all safety, input, and communication delay constraints.

\begin{remark}
    Since $\bphi{xx}$ is the closed-loop response that maps disturbances $\mb{w}$ to the state $\mb{x}$, the condition that $\Phi^{xx,\,ij}_{(t,\tau)} = 0$ either for all $i>j$ or for all $i<j$ essentially enforces a unidirectional disturbance propagation structure in the closed loop, whereby disturbances affecting one agent can affect its downstream neighbors, but not upstream. Such a requirement is common in applications such as connected vehicles, where disturbances affecting trailing vehicles do not influence the states of those ahead. 
\end{remark}

\section{Numerical Evaluation}
\label{sec:simulation}

To evaluate the performance of the proposed minimal communication controller, we consider a multi-agent system composed of $N$ vehicles, each modeled as a two-dimensional double integrator (dropping the agent indexing $i$ for brevity of notation): 
$\ddot{p}_x = u_x$, $\ddot{p}_y = u_y$, 
where $(p_x, p_y)$ represent the $(x, y)$ position of the vehicle 
driven by the force $(u_x, u_y)$. The state of each vehicle 
is  \scalebox{0.7}{$x = \begin{bmatrix} p_x & p_y & \dot{p}_x & \dot{p}_y \end{bmatrix}^\top$}. We discretize the vehicle dynamics with unit step over a finite time horizon $T = 10$. For different tasks, we shall specify different measurement neighbors $\mathcal{O}_i$ and measurement parameter matrices $[C]^{ij}_t$ for all $i,\,j\in \N$ and $t\leq 10$. 
We add independently and identically distributed uniform disturbance $w_t$ and noise $v_t$ to each agent. 
Unless otherwise specified, we set all coordinates of the disturbances $w_t$ to be in range $[-0.05, 0.05]$ and $v_t $ to be in range $ [-0.05, 0.05]$ for all agents. For all experiments, the global control actions are constrained to be within $u\in [-2, 2]^{2N}$.

To validate the effectiveness of the proposed method in reducing inter-agent communication overhead, we compare it against 2 benchmark controllers:
\begin{enumerate}
    \item \baseline: We apply the minimal-communication controller for single-agent systems proposed in \cite{aspeel2023low} to the multi-agent setting. \baseline{} minimizes the number of messages transmitted from sensors located at all agents to actuators located at all agents, without considering whether the message is an inter-agent message. That is, it also tries to minimize an agent's use of its own sensory information.
    \item \decentral: We consider the fully decentralized controller where no inter-agent communication is allowed. The controller is synthesized by searching for any feasible SLS controller such that $ \bphi{*,\,ij} = 0$ for all $*\in \{xx, xy,ux,uy\}$ and $i\neq j$.
\end{enumerate}

To minimize the rank of the SLS parameters as described in \eqref{eq:sls-rank}, we use the nuclear norm as a convex relaxation. To make the optimization process of nuclear norm more numerically stable, the reweighted nuclear norm minimization method is applied~\cite{Mohan2010Reweighted}.

\subsection{Distance tracking under asymmetric control and noise}
\label{section: Two agents among which one is with large noise} 

In this experiment, we consider two vehicles where Vehicle 1 has very little control authority and experiences significantly larger state disturbances than Vehicle 2. In particular, for the global disturbance vector $w_t \in \mathbb{R}^8$ where the first four coordinates correspond to Vehicle 1, we set $w_t \in [-0.25,0.25]\times[-0.60,0.60]\times [-0.05,0.05]^3 \times  [-0.10,0.10]\times[-0.05,0.05]^2 $. We also set the global measurement noise vector $v_t\in [-0.05,0.05]^4$, where the first two coordinates correspond to Vehicle 1. Simultaneously, we equip Vehicle 2 with more control authority than Vehicle 1. 
The safe control task here is for the two vehicles to start and end at specified positions, while tracking their relative distance to each other. In particular, the distance between the two agents is required to satisfy the $L_1$-distance at specific time steps defined as 
\begin{equation}
\label{eq:distance}
\|x^i_t(1:2)-x^j_t(1:2)\|_{1} \leq d_t(i,j),
\end{equation}
where $x^i_t(1:2)$ denotes the first 2 coordinates of Vehicle $i$'s state at time $t$ and $d_t(i,j) >0$ is the specified distance requirement at time $t$. We consider $d_3(1,2) = d_7(1,2) = 5$ and {$d_t(1,2) = 12$ for all $t\in \{1,\ldots,10\}$ in this experiment. 


 In \Cref{tab:compact_message_table}, we report the number of inter-agent messages required by the benchmark controllers and the proposed method. It is evident that the proposed controller significantly reduces the number of messages transmitted between the two agents compared to \baseline. On the other hand, due to the noise and control asymmetry, in order to satisfy the control task, Vehicle 2 must gather information from Vehicle 1 through communication. Therefore, \decentral{} is infeasible in this case. 

 In \Cref{fig: 2uwv_T=10_noisy}, we plot two runs, visualized with two trajectories, of this experiment using the proposed controller with different initial positions. Interestingly, to ensure safe control using only a single message, the proposed controller adopts a proactive strategy. In particular, Vehicle 2 deliberately overshoots the distance-tracking requirement to preempt the need for future communication, anticipating the large disturbances affecting Vehicle 1.

\begin{figure}[h]
    \centering
    \includegraphics[width=0.6\linewidth]{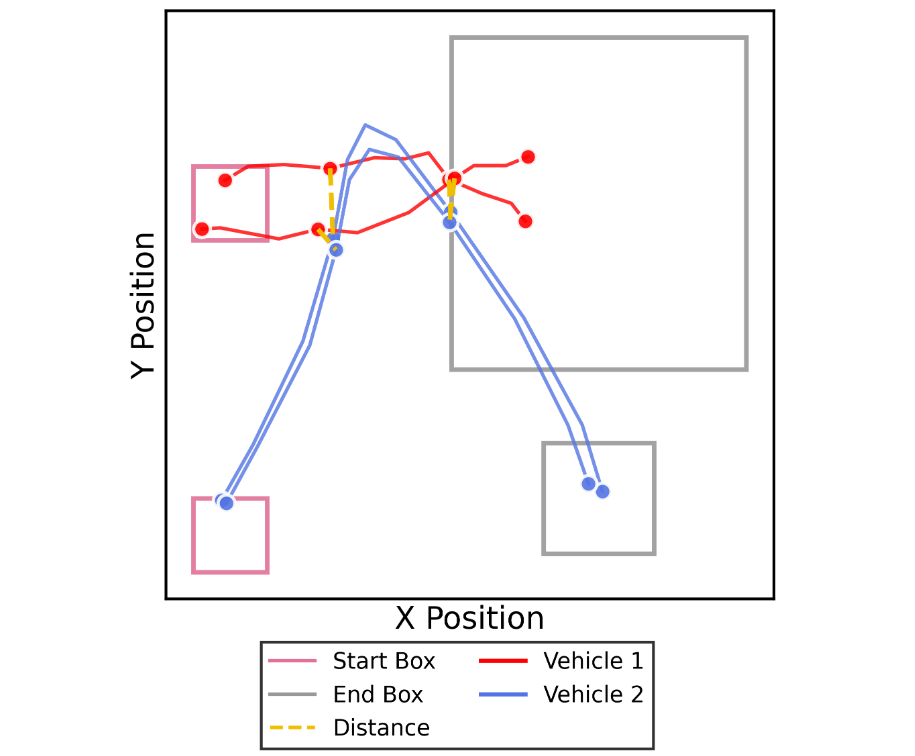}
    \caption{Two sets of trajectories generated by proposed controller for the experiment with asymmetric control and noise.}
    \label{fig: 2uwv_T=10_noisy}
\end{figure}
\subsection{Trajectory tracking with various sensing scenarios}
\label{section: Two agents where one with relative observation}
In this experiment, we consider two vehicles where each vehicle has its own task of passing through a middle waypoint (indicated with a box) at time $t= 5$. Under this task, we consider three different sensing scenarios that are common in practice:
\begin{itemize}
    \item Decoupled measurements: Vehicle 1 and 2 can observe their own positions and communicate with each other.
    \item Relative measurements: Vehicle 1 is equipped with accurate positional devices, while Vehicle 2 can only measure the relative positional information relative to the first vehicle.  
    \item Heterogeneous sensors: Vehicle 1 is equipped with sensors that can measure lateral positions, i.e., the $x$ positions. Therefore, Vehicle 1 can measure the $x$-axis positions of both vehicles. On the other hand, Vehicle 2 can only measure the $y$-axis positions for both vehicles.
\end{itemize}
The two vehicles are required to maintain an $L_1$ distance of $d_t(1,2) = 15$ for decoupled measurements, $15$ for relative measurements, and $14$ for heterogeneous sensors experiments, for all $t \in {1, \ldots, 10}$.


In \Cref{fig: simulation_two_combined}, we illustrate four sets of trajectories generated by the proposed method with different initial positions for the relative measurements and heterogeneous sensors experiment. We report the number of inter-agent messages required by the benchmark controllers and the proposed method in \Cref{tab:compact_message_table} for all three cases. Similar to the previous experiment, the proposed method requires the least number of inter-agent messages compared to benchmarks.

To understand the effect of communication delay on the proposed controller, we impose increasing uniform communication delays on 
the experiment with heterogeneous sensors. The result is shown in \Cref{fig: delay_num_message}. Indeed, as the amount of delay in inter-agent communication increases, the amount of transmission also increases. We note that the proposed method seems more sensitive to communication delay than \baseline. It will be interesting to understand fundamental limitations imposed by communication delays on minimal inter-agent communication control design. 

\begin{figure}[h]
    \centering
    \includegraphics[width=0.85\linewidth]{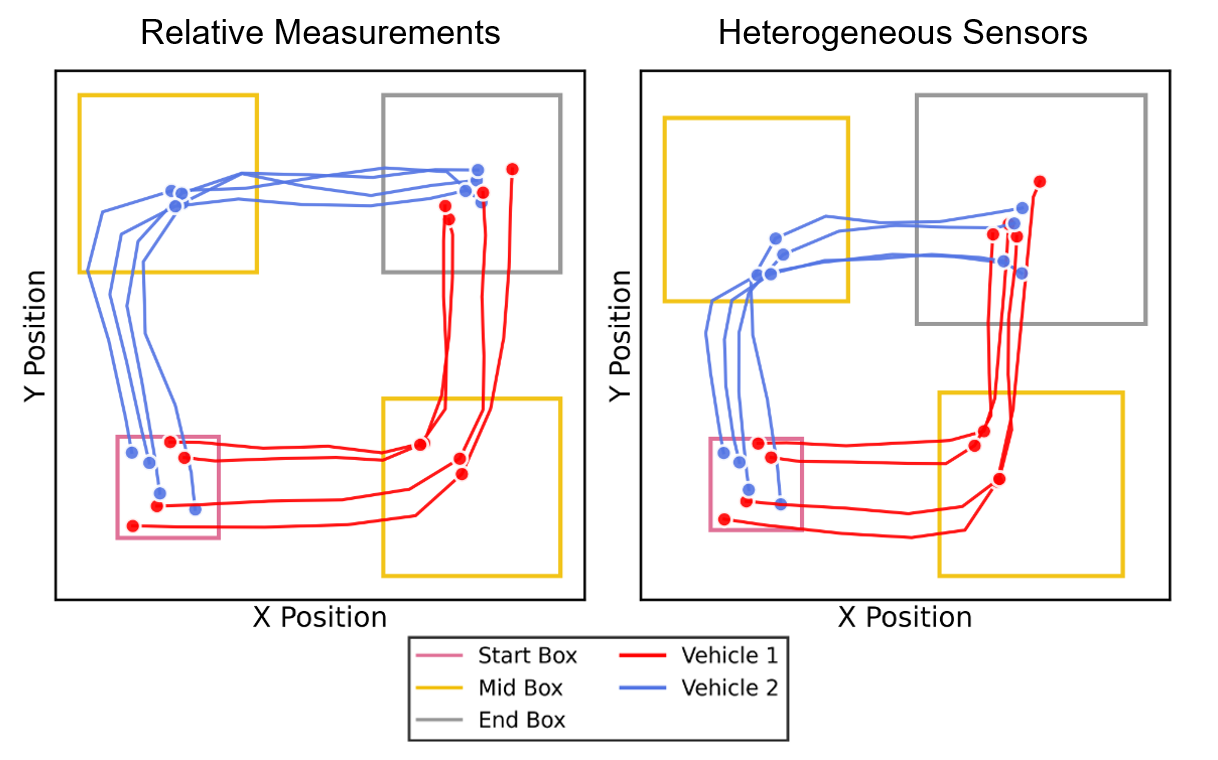}
    \caption{Four sets of trajectories generated by proposed controller in experiments of two vehicles with different sensing strategies.}
    \label{fig: simulation_two_combined}
\end{figure}


\begin{figure}[h]
    \centering
    \includegraphics[width=0.52\linewidth]{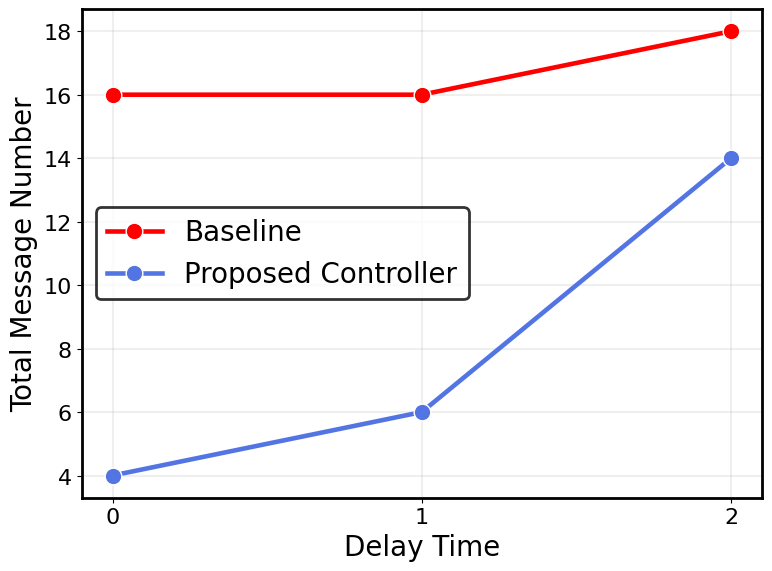}
    \caption{Relationship between delay and the number of message for the baseline and proposed controller in experiment with heterogeneous sensing vehicles.}
    \label{fig: delay_num_message}
\end{figure}

\subsection{Four vehicles chasing  with relative measurement}
\label{section: Four agents with relative measurement}

In this experiment, we consider four vehicles with relative measurement, where Vehicle 1 is equipped with accurate sensors that provide absolute position measurements. On the other hand, Vehicle 2 and 4 can only measure their relative position with respect to Vehicle 1, while Vehicle 3 can only measure its position relative to Vehicle 2. All vehicles are required to pass through a middle box-shaped waypoint at $t=5$ and achieve a counter-clockwise trajectory along way points. 

We visualize the experiment in \Cref{fig: 4uwv_T=10} where we plot four different runs using the proposed controller with different initial positions.
Take Vehicle 1 (pink) in Figure~\ref{fig: 4uwv_T=10} as example, starting from the lower-left box, it is required to pass the yellow middle box at $t=5$ and arrive at the upper-right pink goal box at $t=10$. During the process, all vehicles are required to stay close to others. This constraint is effectively equivalent to a coordinated chasing task among the four vehicles.  Specifically, $d_t(1,2) = d_t(2,3)= d_t(3,4) = d_t(4,1) = 10 \text{ for all } t\in \{1,...,10\}$. 
The resulting number of inter-agent messages for different controllers can be found in \Cref{tab:compact_message_table}. In particular, the proposed method uses $53\%$ less inter-agent transmissions compared to the baseline.  





\begin{figure}[h]
    \centering
    \includegraphics[width=0.8\linewidth]{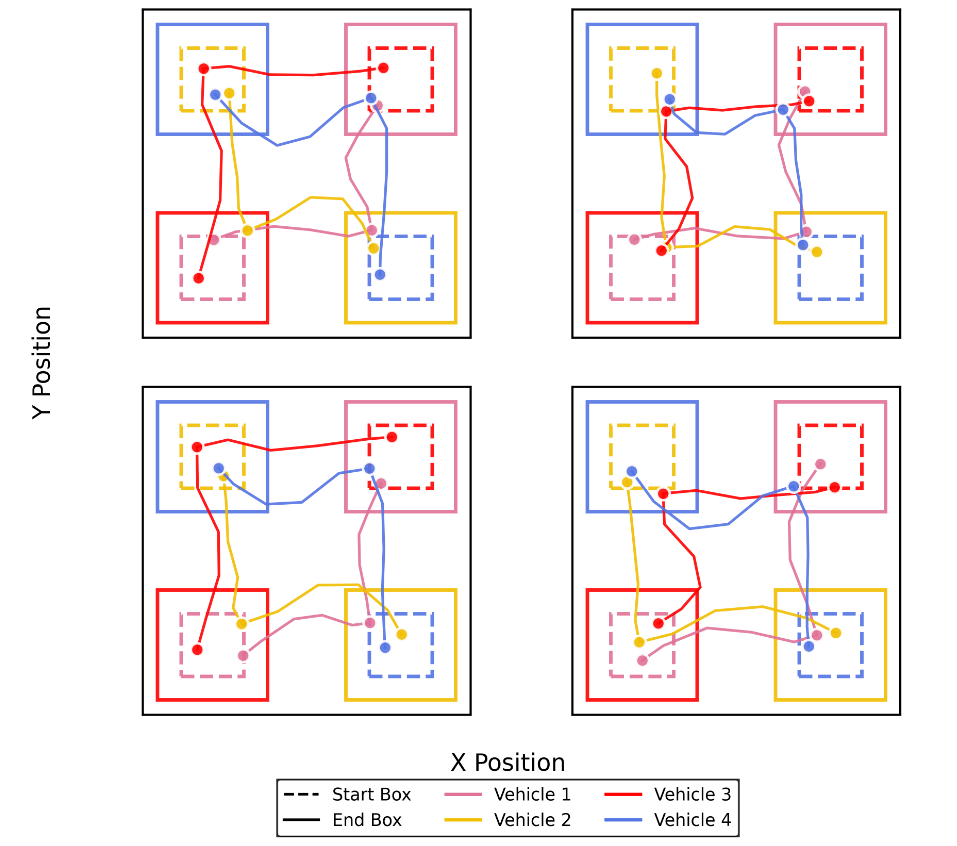}
    \caption{Four sets of trajectories generated by proposed controller designed for four vehicles with relative measurements. Different colors of start and end boxes correspond to the vehicle trajectories with the same color.}
    \label{fig: 4uwv_T=10}
\end{figure}

\begin{table}[ht]
\centering
\renewcommand{\arraystretch}{1.1}
\resizebox{0.48\textwidth}{!}{
\begin{tabular}{llcccc}
\toprule
\textbf{Task} \textbf{(\# Message)} & \textbf{\baseline} & \textbf{\decentral} & \textbf{Ours} \\
\midrule
\multicolumn{4}{l}{\ref{section: Two agents among which one is with large noise}: Asymmetric Control and Noise} \\
& $\boldsymbol{21}$ (12,9)  & $\boldsymbol{-}$ & $\boldsymbol{1}$ (1,0) \\
\midrule
\multicolumn{4}{l}{\ref{section: Two agents where one with relative observation}: Decoupled Measurements} \\
&  $\boldsymbol{16}$ (8,8) & $\boldsymbol{0}$  & $\boldsymbol{0}$ \\
\midrule
\multicolumn{4}{l}{\ref{section: Two agents where one with relative observation}: Relative Measurements} \\
&  $\boldsymbol{25}$ (14,11) & $\boldsymbol{-}$  & $\boldsymbol{12}$ (12,0) \\
\midrule
\multicolumn{4}{l}{\ref{section: Two agents where one with relative observation}: Heterogeneous Sensors} \\
{delay = 0} & $\boldsymbol{16}$ (8,8) & $\boldsymbol{-}$ & $\boldsymbol{4}$ (2,2)\\
{delay = 1} & $\boldsymbol{16}$ (8,8) & $\boldsymbol{-}$ & $\boldsymbol{4}$ (3,3)\\
{delay = 2} & $\boldsymbol{18}$ (9,9) & $\boldsymbol{-}$ & $\boldsymbol{14}$ (7,7)\\
\midrule
\multicolumn{4}{l}{\ref{section: Four agents with relative measurement}: Four Vehicles} \\
& $\boldsymbol{94}$& $\boldsymbol{-}$ & $\boldsymbol{44}$\\
\bottomrule
\end{tabular}
}
\caption{Inter-agent message counts under different control strategies. The bold number indicates the total inter-agent messages sent, while the numbers in parentheses indicate the messages sent from Vehicle 1 to 2 and from Vehicle 2 to 1, respectively. The sign "-" means the controller is not feasible for that task. 
}
\label{tab:compact_message_table}
\end{table}

\section{Conclusion}
This paper proposes an algorithm to design controllers that minimize inter-agent communication for multi-agent systems while ensuring the satisfaction of coordination, safety, and communication delay constraints. 
By formulating the control design problem as a rank minimization problem and employing the System Level Synthesis framework, we derive a tractable convex relaxation. Simulation results demonstrate that our method achieves substantial reductions in inter-agent transmissions compared to existing benchmarks. A current limitation is that our theoretical analysis presently covers only two-agent systems. Future research will focus on extending the theoretical analysis to general multi-agent networks and investigating the fundamental impact of communication delays on minimal-communication control.

\vspace{5pt}
\noindent
\textbf{Acknowledgments:}  
We would like to thank Dr. Kyle Crandall from US Navy Research Lab for helpful discussions motivating several of the examples and the problem formulation.

\bibliographystyle{IEEEtran}
\bibliography{references}

\appendix
\label{appendix}

\textbf{Notation. } We say a matrix $N\in\mathbb{R}^{n\times n}$ is nilpotent if there exists $k\in\mathbb{N}_+$ and $k\leq n$ such that $N^k = 0$. We use binary matrices to denote sparsity patterns. For two binary matrices $\mathcal{S}_1$ and $\mathcal{S}_2$, the operation $\mathcal{S}_1 + \mathcal{S}_2$ performs an element-wise OR. Let $\spar{M}$ denote the sparsity of a matrix $M\in\mathbb{R}^{m\times n}$ where $\spar{M} \in \{0,1\}^{m\times n}$. We say ${M} \in \mathcal{S}$ if $\spar{M}+ \mathcal{S} = \mathcal{S}$. Furthermore, we will define three special sparsity patterns for $\bphi{*}$. In particular, we define the following sparsity patterns for block lower triangular matrices with $T$ by $T$ many sub-blocks:
\begin{itemize}
    \item Pattern $\D$: all sub-blocks are composed of block diagonal mini-blocks
    \item Pattern $\U$: the $(1,2)$th mini-block inside all sub-blocks are non-zero and zero everywhere else
    \item Pattern $\Lo$: the $(2,1)$th mini-block inside all sub-blocks are non-zero and zero everywhere else,
\end{itemize}
where for any $*\in\{xx, xy,ux,uy\}$, the sub-blocks of $\bphi{*}$ are $\Phi^{*}_{(t,\tau)}$ for all $t,\,\tau \in \{0,\ldots,T\}$ and $t\leq \tau$. The mini-blocks inside the sub-blocks of $\bphi{*}$ are $\Phi^{*,ij}_{(t,\tau)}$ for $i,j \in \{1,2\}$. 
For simplicity, we sometimes overload the notation and use $\mathcal{S}\left(M\right)$ for a binary matrix $\mathcal{S}$ and a matrix $M$ to mean the application of a masking operation of $\mathcal{S}$ on $M$, where $\mathcal{S}\left(M\right)(i,j) = 0$ if $ \mathcal{S}(i,j) = 0$ and $\mathcal{S}\left(M\right)(i,j) = M(i,j)$ otherwise.

\textbf{Simple matrix facts. } For the rest of the appendix, we will use the fact that adding zero rows and columns to a matrix does not change its rank. We will also invoke the following inequalities: F1) $\text{rank}(A+B)\leq \text{rank}(A) + \text{rank}(B) $, F2) $\text{rank}(AB)\leq \min\{\text{rank}(A) ,\, \text{rank}(B)\} $, and F3) $\text{rank}\left(\sum_{k=1}^{n} \alpha_k A^k \right) \leq \text{rank}(A)$ for all $\alpha_k \in \mathbb{R}$.

\subsection{Proof of \Cref{lem:phi-rank}}

First, note that if for all $\tau,\,t \in \{ 0,\, \ldots, T\}$, $\Phi^{xx,\,12}_{(t,\tau)} = 0$, then $\bphi{xx} \in (\D + \Lo)$. 
Without loss of generality, we assume this is the sparsity pattern of $\bphi{xx}$. The case of $\bphi{xx} \in (\D + \U)$ can be handled analogously.

To show $\text{rank}(\mb{K}^{21}) \leq \!\!\underset{*\in \{xx,xy,ux,uy\}}{\sum} \!\! \text{rank}(\bphi{*,\,21})$, we will show that $\mb{K}^{21}$ is composed of linear combination of $\Lo\left(\left(\bphi{xx}\right)^{-1}\right)$ and $\bphi{*,21}$ for $*\in \{ xy,ux,uy\}$. In particular, due to \eqref{eq:K-phi}, it is clear that $\bphi{uy,21}$ contributes an additive component to $\mb{K}^{21}$. 

We now consider the term $\bphi{ux} \left(\bphi{xx}\right)^{-1} \bphi{xy}$. First, we claim that $\left(\bphi{xx}\right)^{-1} \in (\D + \Lo)$. To see this, note that due to the affine subspace constraint \eqref{eq:sls}, $\bphi{xx}$ is required to have $ \Phi^{xx}_{(t,t)} = I$ for all $t\in\{0,\ldots,T\}$. Therefore, $\bphi{xx} = I + N$ with $I \in \mathbb{R}^{(T+1)n_x \times (T+1)n_x}$ and $N$ a block strictly lower triangular matrix, which is nilpotent. In particular, there exists $n\leq T+1$ such that $N^n = 0$. 
Therefore, we can write $\left(\bphi{xx}\right)^{-1} = I + \sum_{k=1}^{n-1} (-1)^k N^k$. Since multiplication and summation of block lower triangular matrices remain block lower triangular, $\left(\bphi{xx}\right)^{-1} \in (\D + \Lo)$. 
Furthermore, note that $\bphi{xx,21}$ corresponds to the entries of $\bphi{xx}$ supported on the sparsity pattern defined by $\Lo$. Referring to \Cref{tab:sparsity}, which can be verified via simple calculation, we see that if $A\in\D$ and $B\in \Lo$, then $A^2 \in \D$, $B^2 = 0$, and $AB \in \Lo$. Therefore, for all $k\in \{1,\ldots,n\}$, the entries of $N^k$ supported on the sparsity pattern defined by $\Lo$ are all linear combination of summation of multiplication of $\bphi{xx,21}$. Consequently, the same holds for $\left(\bphi{xx}\right)^{-1}$. Using F1 and F3, the rank of $\Lo\left(\left(\bphi{xx}\right)^{-1}\right)$ is at most $\text{rank}(\bphi{xx,21})$.

For each one of $\bphi{ux},\, \left(\bphi{xx}\right)^{-1}$, and $ \bphi{xy}$, we apply a masking operation using the three binary matrices $\D,\, \U,\,$ and $\Lo$. This operation decomposes each matrix in $\{\bphi{ux},\, \left(\bphi{xx}\right)^{-1},\,\bphi{xy}\}$ into three additive components, corresponding to the sparsity patterns defined by the masks. In particular, the component matrices that are constructed with $\Lo$ corresponds to the entries of $\bphi{ux,21},\, \Lo\left(\left(\bphi{xx}\right)^{-1}\right)$, and $ \bphi{xy,21}$, while the component matrices that are constructed with $\U$ corresponds to the entries of $\bphi{ux,12}$ and $ \bphi{xy,12}$. 
By examining the outcome of the all nine possible multiplications of the component matrices for the operation of $\bphi{ux} \left(\bphi{xx}\right)^{-1} \bphi{xy}$, which is listed in \Cref{tab:sparsity}, we see that indeed $\mb{K}^{21}$ is entirely made up of linear combination of $\bphi{uy,21},\, \bphi{ux,21},\, \Lo\left(\left(\bphi{xx}\right)^{-1}\right)$, and $ \bphi{xy,21}$. 
Therefore, using F1 and F2, we have $\text{rank}\left(\mb{K}^{21}\right) \leq {\sum}_{*\in \{xx,xy,ux,uy\}}\text{rank}(\bphi{*,\,21})$. 
The same argument above holds true for $\mb{K}^{12}$. This concludes the proof.

\hfill $\square$

\begin{table}[h!]
\centering
\renewcommand{\arraystretch}{1.1} 
\begin{tabular}{|c|c|c|c|}
\hline
\textbf{} & $\D$ & $\Lo$ & $\U$ \\
\hline
$\D$ & $\D$ & $\Lo$ & $\U$ \\
\hline
$\Lo$ & $\Lo$ & 0 & $\D$ \\
\hline
$\U$ & $\U$ & $\D$ & 0 \\
\hline
\end{tabular}
\caption{Resulting sparsity structure from the product of two sparse matrices.}
\label{tab:sparsity}
\end{table}


\end{document}